# Role-based Label Propagation Algorithm for Community Detection

Xuegang Hu, Wei He[*], Huizong Li, Jianhan Pan

School of Computer & Information, Hefei University of Technology, Hefei, 230009, China

## Abstract

Community structure of networks provides comprehensive insight into their organizational structure and functional behavior. LPA is one of the most commonly adopted community detection algorithms with nearly linear time complexity. But it suffers from poor stability and occurrence of monster community due to the introduced randomize. We note that different community-oriented node roles impact the label propagation in different ways. In this paper, we propose a role-based label propagation algorithm (roLPA), in which the heuristics with regard to community-oriented node role were used. We have evaluated the proposed algorithm on both real and artificial networks. The result shows that roLPA is comparable to the state-of-the-art community detection algorithms.

Keywords: Community-oriented; node role; heuristics; label propagation; community detection; complex networks.

## 1 Introduction

Many real-world systems can be modeled as network, such as on-line social networks, scientist collaboration networks, epidemic networks, electric networks, the Internet, World Wide Web, and metabolic networks. Community structure [1] of network is the tendency for nodes to be assembled into groups, or communities, which densely connected inside and loosely connected with the rest of the network. For many real-world systems, community structure provides comprehensive insight into their organizational structure and functional behavior. For example, spreading of epidemic disease, which can be seen as a dynamic process on network, is greatly affected by the community structure of the social network [2].

Community detection algorithms, which aim to reveal hidden community structure in network, have attracted significant attention in recent few years. A substantial number of community detection algorithms have been proposed, including modularity optimization algorithms, spectral clustering algorithms, hierarchical partition algorithms, and information theory based algorithms [3]. Among them, the label propagation algorithm (LPA) [4] proposed by Raghavan et al. is one of the most commonly adopted algorithms with nearly linear time complexity. Owing to its prominent speed, conceptual simplicity, accurateness, and easy implementation of parallelism, LPA is suitable for large-scale networks with millions of nodes and edges, e.g. Facebook, which had almost 1.4 billion monthly active users by the end of 2014.

Despite various advantages, some issues of the LPA have not been properly addressed, such as the deficiency of robustness and stability due to its random updating and breaking tie strategy. The weak robustness of the LPA means the solutions in different runs can be quite different. And that is related to the significance of community structure in a network, that is, for a network with weaker community structure, the LPA would get the less stable solution. Also for a network with weak community structure, the LPA may produce a monster community, or giant community, which dominate a big part of the network and swallow the small communities, making the solution trivial [5].

*Corresponding author. Tel.: +86 13866151397.
E-mail address: ahhfhw@163.com (W. He).

Raghavan et al. point out that, if the core nodes of network are identified, the LPA can be implemented by initializing core nodes with unique labels and leaving the other nodes unlabeled [4]. In this case the unlabeled nodes will have a tendency to acquire labels from their closest attractor and join that community. However, this implementation attracted little attention due to its inherent limitation, that is, identifying the core node of community before identifying the community itself is very difficult. On the other hand, identifying the core nodes without considering community structure may lead the algorithm toward an undesired solution. For example, the core nodes identified through the degree centrality may be the inter-community hubs, e.g. overlapping nodes with high degree shared by two or more communities, which would lead the algorithm to failure.

In this paper, we propose a role-based label propagation algorithm (roLPA) by introducing the heuristics inspired by community-oriented node role, and defined some new measure for the node preference by considering the community-oriented role. We apply the algorithm on both real and artificial networks. The algorithm is shown to be comparable to the state-of-the-art community detection algorithms, with nearly linear time complexity.

The rest part of the paper is arranged as follows. In Section 2 we review label propagation algorithm and community-oriented node role. In Section 3 we stated the proposed algorithm. In Section 4, we present the result of experiments. Section 5 gives the summary and discussion of this paper.

## 2 Related Work

### 2.1 LPA

The Label Propagation Algorithm (LPA) was proposed by Raghavan et al. [4], which uses network structure alone, requiring neither a pre-defined objective function nor prior information about the communities.

The main idea behind LPA is to propagate the labels of node throughout the network and form communities through the process of label propagation itself. Intuitively, a single label would be trapped inside a densely connected group of nodes during label propagating. At first, each node is initialized with a unique label denoting the community it belongs to. Then every node updated its label iteratively. At every step of iteration, a node updates its label as most of its adjacent neighbors currently have. After a few iterations, densely connected groups of nodes tend to form consensuses on some particular labels. The algorithm converges until none of the nodes needs to update its label anymore, or the label of each node is just the same as the most frequent label among its neighbors. Then the communities are identified due to the label of nodes.

The algorithm works as follows:

A networks is described as a graph $G(V, E)$, where $V$ is the set of nodes and $E$ is the set of edges. For node $i$ ($i \in V$), let $L_i$ denote the label of $i$, $N(i)$ denote the set of its neighbors. At the beginning, each node is initialized with a unique label, e.g. $L_i = i$. At every step of iterations, each node updates its label to the one shared by the most of its neighbors,

$$L_i = \arg\max_l |N^l(i)| \tag{1}$$

where $N^l(i)$ denotes the set of neighbors of node *i* that have the label l, and $|X|$ is the cardinality of set *X*. If more than one label are the most frequent ones, a new label is chosen randomly from them. To avoid fluctuations, the current label is kept if it is one of the most frequent ones in some implementation. The update-order of nodes is also random in iterations. In the process of label propagation, densely connected nodes tend to reach a consensus on a unique label quickly.

The label propagation is performed iteratively until each node has (one of) the most frequent label of its neighbors, that is, none of the nodes need to changes its label. Finally, communities are identified as groups of nodes sharing the same labels. For most situations, Raghavan et al. claim that 95% of nodes are already accurately clustered after 5 iterations, and the algorithm would converge after no more than 10 iterations. So the number of iterations to converge appears independent of the size of network. The LPA runs linearly to the number of edges with time complexity O(*km*), where *k* is the number of iterations and *m* is the number of edges.

Due to the introduced randomize, e.g. random update order and the random tie breaking strategy, the algorithm produces different solutions at different runs. This issue, or robustness of the algorithm, is related to the significance of community structure in a network [6].

For the networks with strong community structure, the different solutions are usually similar to each other. You can run the algorithm for many times and aggregate all the solutions to get one solution with most useful information. For those networks with weak community structure, the result derived by aggregating different solutions, one of which can differs a lot from the other, may be trivial.

Leung et al. first introduced heuristics know as hop attenuation and node preference to label propagation algorithm [5]. Barber et al. reformulated the LPA as a modularity optimization problem and introduced the modularity-specialized LPA (LPAm) [7] after Tibély et al. have shown that label propagation is equivalent to a large zero-temperature kinetic Potts model [8]. Liu et al. combined the LPAm with a multistep greedy agglomeration for escaping from poor local optimum [9]. Šubelj et al. used a node balancer as the node preference that can counteract for the introduced randomness [10]. Another advanced LPA presented by Šubelj et al. use the defensive diffusion and attenuation to extract the cores and whisker communities of network [11]. Ugander et al. presented the problem of balanced label propagation partitioning, which partition a graph into k balanced parts using the label propagation method later [12].

Generally speaking, one obtains advanced LPA through four ways:

1. Introducing the hop attenuation, which decreases while the label traverse, in order to prevent the occurrence of monster community. As a result, the solution is susceptible to the parameter, and it is difficult to determine the proper parameter.

2. In order to improve the accuracy of the LAP, some limitation can be adopted, e.g. modularity, to significantly reduce the solution space of the equivalence optimization problem. Usually, the running time and space complexity increases significantly. On the other side, the objective function of optimization problem, such as modularity, may lead to locally optimal solution of the whole solution space.

3. Some algorithms use node preference to counteract the introduced randomness. For all of these algorithms, the heuristic definitions of node preference have little correlation with the community structure of networks, which may lead to poor solution. In addition, some of them have high complexity, which significantly weaken the speed advantage of the LPA.

4. The algorithm is sensitive to the updating order of nodes in each step, thus the updating order can also be predefined based on some measurement to counteract the introduced randomness. This method can enhance the stability of the algorithm. But there is no adequate measurement considering the community structure of the network by now.

## 2.2 Community-oriented Node Role

In complex network analysis field, node roles represent node-level connectivity patterns such as nodes that bridge different regions, and identifying the role of a node is one of the key issues of network analysis. There had been many metrics that can be used to assign roles to individual nodes, such as betweenness, degree and closeness. But most of these metrics do not take into account the community structure of network.

In 2007, Scripps et al. first defined four types of community-oriented roles (e.g. ambassadors, big fish, loners, and bridges) according to the number of communities and links incident to it [13]. According to the definition, the ambassadors provide connections to many different communities with a high degree and a high community score. The big fish are very important only within a community with a high degree but a relatively small community score. The Bridges serves as bridges between a small number of communities with a low degree but a high community score. Finally, the loners have a low degree and low community score. They also proposed a method named rawComm to identify these roles basing on the cliques of the network. After showing how existing link mining techniques can be improved by knowledge of such roles, they also proposed a community-oriented metric for estimating the number of communities linked to a node.

Chou et al. (2010) pointed out that assigning roles to nodes connected to communities with few links by rawComm is difficult, due to the proportional feature of the degree and the community metric [14]. They divided the nodes between different communities into three types of role -- bridge, gateway and hub. A bridge is located between two communities, connecting communities, each of which has only one link with it. A gateway node acts as an entrance of a community, to which most of neighbors of the gateway node belong. A hub is a confluent node, on which communities converge, when groups of nodes are divided into different communities. Furthermore, they proposed an identifying algorithm requiring no information of the community structure.

After investigating the relation between eigenvalues and the community structure, Wang et al. (2011) suggest a basic approach to define the importance of nodes using the spectrum of the graph [15]. They introduced a method to divide the important nodes into community cores and bridges using the eigenvectors of graph Laplacian.

Based on the current flow in the electrical circuits, Zhu et al. (2014) developed an algorithm to identify the overlapping nodes and bridging nodes without partitioning all communities explicitly [16]. The two types of critical nodes connected to the other nodes within communities with relatively high current flow passing through them, as captured by the centrality of current flow. Further more, they distinguished the two types of nodes from each other via the imbalance of flows along their edges.

To efficiently assess different characteristics of a node, an intuitive idea is to select some indicators with weak correlations. However, Huang et al. (2014) argued that it is much better to select the indicators with strong correlations [17]. As they pointed out, indicator correlation is based on the statistical analysis of a large number of nodes, the particularity of an important node will be outlined if its indicator relationship doesn't comply with the statistical correlation. On the basis of correlation analyses of typical indicators, they proposed the methods EIMI and RUMI to evaluate the importance of nodes and identify the node role based on the selected indicators, that is, degree, ego-betweenness centrality and eigenvector centrality.

Up to now, identifying community-oriented node roles fast and accurately without priori information about community structure is still a very challenging and open problem. Though various kinds of roles could be defined from different perspectives, two basic kinds are widely accepted considering about community structural. The first kind of role plays the bonding role usually takes up the central position of the community, namely core node. The second kind of role connects two or more communities together, namely bridge node.

## 3 Role-based Label Propagation Algorithm

In this paper, we propose the role-based label propagation algorithm (roLPA) by introducing the heuristics concerning with community-oriented roles.

Due to the introduced randomize, the LPA algorithm produces different solutions at different runs. The final solution can be derived from aggregating different solutions together. But for a network with weak community structure, the aggregating result may be trivial, because each one may differ a lot from the others. Also for a network with weak community structure, the LPA may produce a giant community, which dominate a big part of the network and swallow the small communities, making the solution trivial.

The intuition of the roLPA algorithm is to improve label propagation by considering the impact of different community-oriented roles. We analysis various situations of the label propagation for different roles, and employ some heuristics with regard to community-oriented roles. By employing these heuristics, roLPA implement the propagation in two phases, namely balancing propagation and converging propagation. During the balancing propagation, the community core nodes are identified and the bridge nodes are kept unfixed. During the converging phase, the propagation is accelerated to converge.

### 3.1 Intra-community and Inter-community Node Roles

Our method considers node role from two different perspectives, that is, intra-community topology features and inter-community affiliation.

**I. Intra-community Role**

Assuming the communities of a network are already identified, some basic kinds of node role are widely accepted referring to different topology features of a node in a community, e.g.

core, peripheral, and tight-knit participant.

The core nodes, which have low clustering coefficient, high degree and high centrality, take up the central position of communities and bond other nodes together. The peripheral nodes, which usually have low degree and high eccentricity, do not disconnect the graph but often increase geodesic lengths when removed. Core and peripheral constitute the centralized group. The extreme example of the centralized group is star, as Fig 1(left) shows.

In contrast, the tight-knit participant nodes, which have high degree, high clustering coefficient, and high homophily, can be found in tightly coupled groups. The extreme example of tightly coupled group is clique, as Fig 1(right) shows.

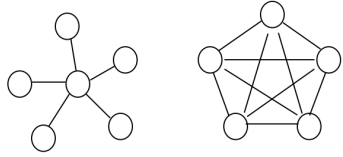

Figure 1. Star (composed with core and peripheral) and tight-knit clique.

These three types of node role are concerned with the topology feature in community, that is, considering the intra-community nodes and edges only. There are many indicators to measure the topological feature of community and role, here we select some of them without detailed comparison. In order to measure the density of community, we introduce the edge density, which is defined as:

$$Density(c) = \frac{2*E(c)}{V(c)*(V(c)-1)} \quad (2)$$

where $E(c)$ is the count of intra-community edges in community $c$, $V(c)$ is the count of nodes in community $c$. The community composed with core and peripheral would have low density, while the tight-knit community would have high density.

The indicators of centrality identify the most important nodes, or cores, within a network. In this paper, we use the intra-community degree centrality due to its locality, which is defined as:

$$Centrality(i) = \frac{D_{in}(i)}{max_{j \in c_j}(D_{in}(j))} \quad (3)$$

where $D_{in}(i)$ is intra-community degree, defined as the number of neighbors of node $i$ within its community. In the community composed with core and peripheral, the centrality of core nodes would be ultra high, while the centrality of peripheral nodes would be low. While in the tight-knit community, there are little different between the centrality of different nodes.

## II. Inter-community Role

In this study we consider only one type of the inter-community role, that is, bridge. A bridge provides connections between different communities. In real world networks, bridge nodes show its importance in exchanging information and resources between different communities.

In order to measure the community-affiliation of nodes, we define the loyalty of node *i* as the ratio of intra-community degree to the degree of the node:

$$Loyalty(i) = \frac{D_{in}(i)}{D(i)} \qquad (4)$$

As the equation indicates, the more intra-community instant neighbors a node has, the higher loyalty it has. Bridge nodes usually have low loyalty, while non-bridge nodes usually have high loyalty.

**3.2 Node Role and Label Propagation**

The LPA algorithm assumes that a single label would be trapped inside a densely connected group of nodes during label propagating. In the different phase of label propagating, various community-oriented roles influence the formation of communities by dissimilar manners. For example, at the beginning phase of the propagation, the core nodes contribute significantly to identifying the community structure by infecting neighbor nodes in the same communities quickly. While the bridge nodes, which connect to many different communities, have negative impact on identifying the community structure because they mixing different communities together. If we have already known the roles of nodes, the solution can be improved by adjusting the label preference and update order. However, identifying the community-oriented node role without community information is a difficult task. On one hand, approximate method may produce poor result. On other hand, precise methods have high time complexity, which significantly weaken the superiority of LPA.

As a consequence, we proposed a two-phase method embedding the node role information into the process of label propagation to produce stable and robust solution.

During the label propagation process, the potential giant community infects other communities in two ways, namely core-first infection and peripheral-first infection. The core-first infection is top-down, that is, the core nodes are infected first, and then the peripheral nodes are infected through the core nodes. On the contrary, the peripheral-fist infection is bottom-up, that is, the peripheral nodes are infected first, and then the core nodes are infected through the peripheral nodes around.

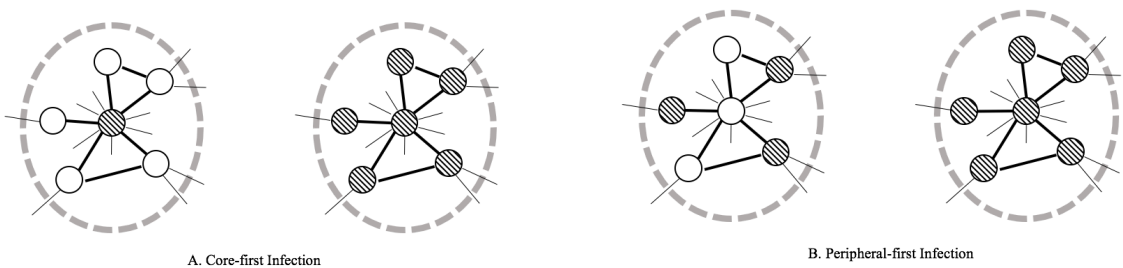

A. Core-first Infection  B. Peripheral-first Infection

Figure 2. Sample network: core-first infection and peripheral-first infection. The color denotes the label of node, and the thick line denotes the intra-community edge.

Due to the topology feature of different roles, core-first infection is usually acute, because the peripheral around it would be infected by the core quickly. Once the core was infected, the peripheral would be infected soon. It usually occurs at the beginning of label propagation, when most adjacent neighbors of the core nodes are non-uniform. Another circumstance occurs when the node is core and bridge at the same time. When the community form consensuses on some particular labels and the core nodes are non-bridge nodes, only peripheral-first infection can arise.

The original LPA updates nodes in a random order. As [10, 18, 19] argued, the LPA algorithm is extremely sensitive to updating order. To study the effect of updating order and inter-community node role, we consider a toy network in Fig 3.

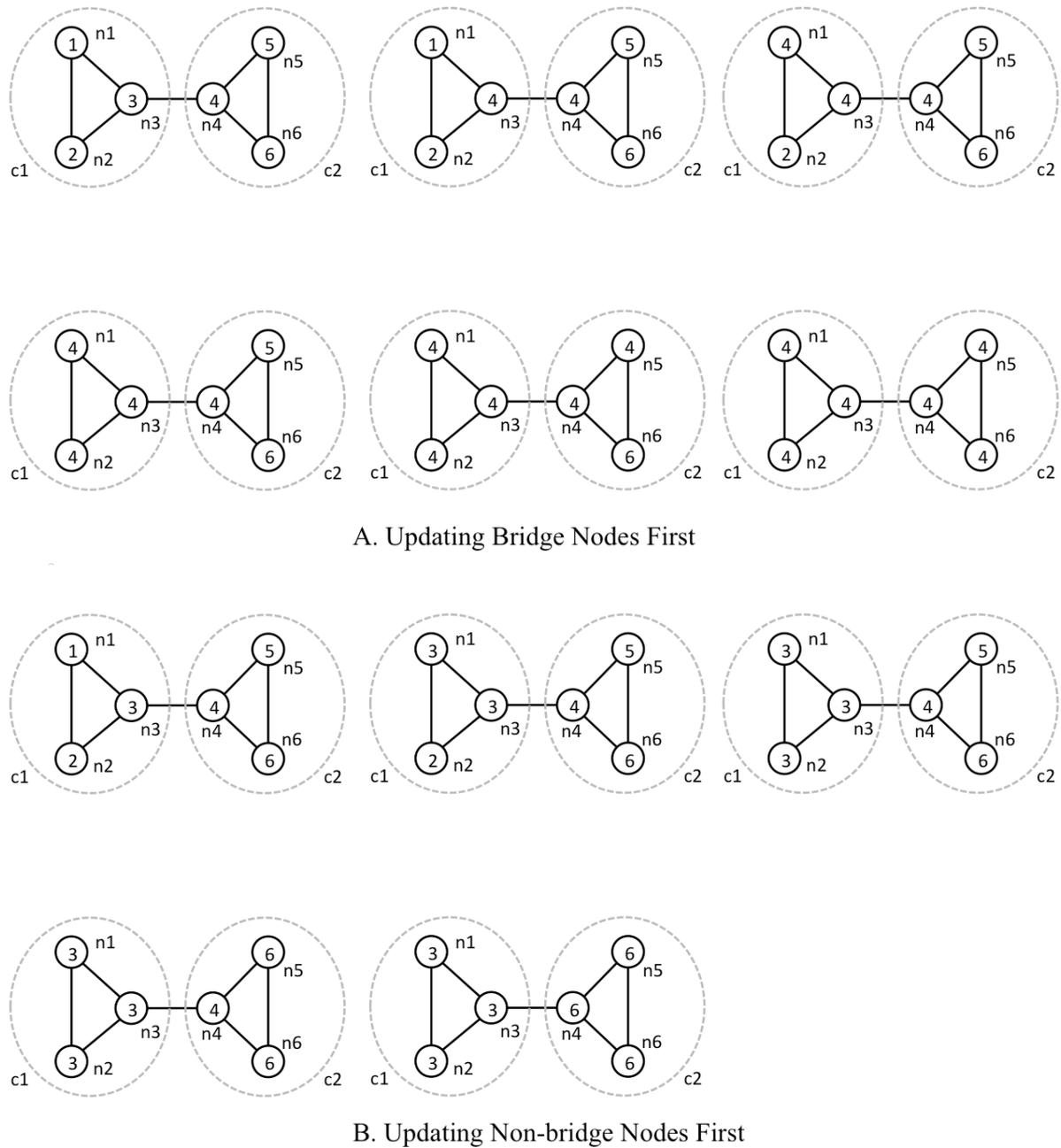

Figure 3. Sample network: updating non-bridge node first (A) and updating bridge node first(B).

The network consists of two communities, namely c1 and c2, where community c1 includes nodes n1, n2, and n3, and community c2 includes nodes n4, n5, and n6. If non-bridge nodes n1, n2, n5, n6 are updated first, the algorithm will lead to the natural community structure of the network. But if we update bridge nodes n3 (or n4) first, it can also adopt the label of node n4 (or n3), then the algorithm will probably identify all nodes as the same community. We can get the conclusion that updating non-bridge nodes first can avoid some core-peripheral infection.

Considering the intra-community node role, we point that the peripheral should be updated first, since this strategy can eliminate the core-first infection effectively. The labels of core nodes have more chance to propagate due to its high degree, so updating the peripheral nodes first can promote them form consensuses with the label of their adjacent core node. The predominance of its label can protect the core nodes from being infected by other nodes, in both formed and unformed community.

As above, by updating the non-bridge nodes before bridge nodes and updating peripheral nodes before core nodes, more stable and precise result can be derived.

### 3.3 Indicator for Update-order

There are more than one indicators that can address the problem of ordering the nodes, however, the detailed comparison is omitted. In our proposed algorithm, Burt's constraint score was used as the ordering indicator due to its locality.

Burt's measure of constraint for node $i$ is defined as

$$Constraint(i) = \sum_{j \in V(i), j \neq i} \sum_{q \in V_i, q \neq i, j} (p_{iq} p_{qj})^2 \tag{5}$$

where $V_i$ is the ego network of node $i$, and the proportional tie strength $p_{ij}$ is defined as

$$p_{ij} = \frac{a_{ij} + a_{ji}}{\sum_{k \in V_i, k \neq i} (a_{ik} + a_{ki})} \tag{6}$$

$a_{ij}$ are elements of $A$ and the latter being the graph adjacency matrix. For undirected network, it can be simplified as

$$p_{ij} = \frac{a_{ij}}{\sum_{k \in V_i, k \neq i} a_{ik}} \tag{7}$$

According to the definition, Burt's constraint is higher if a node has less, or mutually stronger related (i.e. more redundant) neighbors. So a peripheral and non-bridge node would have high constraint score. In our method, we update the nodes in descending order. This strategy makes sure that most of the peripheral and non-bridge nodes are updated before the core or bridge nodes.

### 3.4 Node Preference

Considering node preference and hop attenuation, the updating rule in Eq. (1) rewrites to

$$L_i = \arg\max_l \sum_{j \in N^l(i)} s_j w_j \tag{8}$$

## I. Balancing Propagation

In order to prevent from arising the giant community, the normalized edge density was introduced to the node preference, defined as:

$$w_i = 1 + Density_{norm}(c_i) \tag{9}$$

where $c_i$ is the community of node $i$. The normalized edge density of community $c$ is defined as:

$$Density_{norm}(c) = \frac{Density(c) - min_{c\prime \in C}(Density(c\prime))}{max_{c\prime \in C}(Density(c\prime)) - min_{c\prime \in C}(Density(c\prime))} \tag{10}$$

where $C$ is the set of all communities.

According to the definition, the edge density of the close-knitted group would be high, and the edge density of the sparse group would be low. So this criterion promotes close-knitted communities but restrain sparse ones. As we discussed, the high centralized communities can be formed quickly by updating the peripheral first. By introducing normalized edge density, it would be very difficult for them to swallow other communities due to their comparatively low density. Even the close-knitted communities can't expand quickly, because the density of these communities would sharply decrease after swallowing other sparse communities.

Then we introducing minus normalized loyalty to keep the bridge nodes unfixed and avoid the peripheral-core infection:

$$w_i = 1 + Density_{norm}(c_i) - Loyalty_{norm}(i) \tag{11}$$

The normalized loyalty of node $i$ is defined as:

$$Loyalty_{norm}(i) = \frac{Loyalty(i) - min_{j \in V}(Loyalty(j))}{max_{j \in V}(Loyalty(j)) - min_{j \in V}(Loyalty(j))} \tag{12}$$

where $V$ is the set of all nodes in network.

In a sense, the bridge node between two of more communities can be seen as overlapping parts of these communities. For the original LPA, the influence of a community would increase accumulatively by gradually infecting the bridge nodes around. At the beginning, the infected bridge nodes are usually the peripheral nodes of other communities which have less intra-community neighbors. At last, the core nodes would be infected if enough peripheral nodes are infected. This problem can be avoided by keeping the bridge nodes unlabeled. But this strategy is difficult to implement and may produce new problems. Firstly, it's difficult to identifying the bridge nodes before identifying the community structural. Secondly, keeping some nodes unlabeled would impact their neighbors, and may change the propagation process significantly. To limit the impact of bridge nodes, minus normalized loyalty was introduced for the node preference. This strategy make the bridge nodes alternate their labels between the labels of adjacent communities.

To study the effect of introduced minus normalized loyalty, we consider a toy network in Fig 4(a). The network consists of two communities, namely c1 and c2. Community c1 includes three non-bridge nodes (n1, n2, and n3) and one bridge node (n4). Community c2 includes three non-bridge nodes (n5, n6, and n7). At first node n4 is labeled as c1, the loyalty of n3 is 1, and the loyalty of n5 is 2/3. During the label propagation, node n4 is labeled as c2 due to the higher node preference of n5. Now as the Fig 4(b) shows, the loyalty of n3 decrease to 2/3, and the loyalty of n5 increase to 1. Then at the next iteration of propagation, node n4 will be labeled as c1 again. This alternating process repeats until the balancing propagation ends.

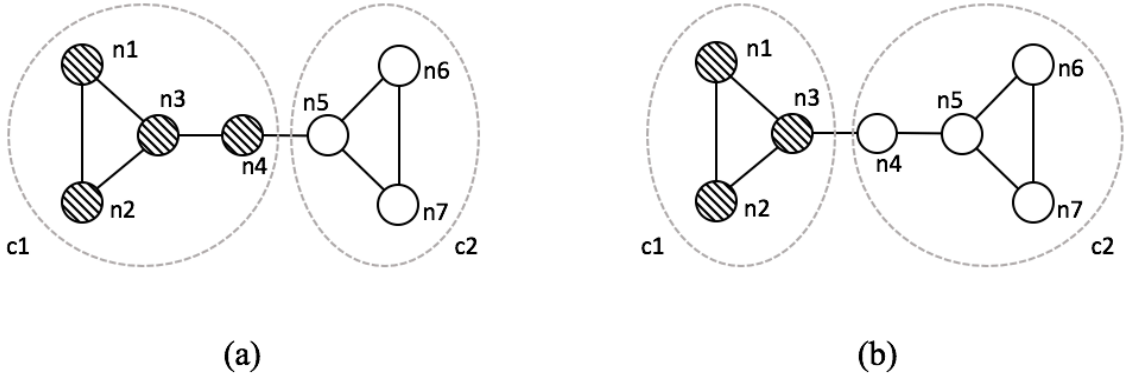

Figure 4. Sample network: balancing propagation.

The balance propagation cannot converge, so we proposed a dynamic strategy as the terminal condition. The strategy is based on the oscillation of changed labels during the iteration. As the potential communities gradually formed, less and less nodes changed their labels, making the count of changed labels in each iteration decrease progressively at the beginning. When most of the potential communities formed, only some boundary nodes would alternate their labels between the labels of adjacent communities, making the count of changed labels in iterations oscillate. Before the final balanced oscillation stage, the depression and oscillation mixed together. For the larger network, there would be more oscillation before balance. Thus we can use the count of oscillation as the approximate indication for balanced stage. We define the upper limit of oscillation count as:

$$Oscillation_{max} = log_{10} n \quad (13)$$

where *n* is the number of nodes in network. For example, the upper limit of oscillation for a network with 1000 nodes is 3, so we terminate the balance propagation when the count of oscillation is up to 3.

As discussed above, the balancing propagation promotes core-peripheral communities and close-knitted communities, but leaves the bridge nodes unfixed.

**II. Converging Propagation**

During converging propagation, our strategy is giving more influence to the non-bridge core nodes of communities, which can reduce the introduced randomness and speedup the converging process. We introduce normalized loyalty and intra-community degree centrality for implementation. The non-bridge nodes have high normalized loyalty, and the core nodes have high centrality. The weight of node $i$ is defined as:

$$w_i = 1 + Loyalty_{norm}(i) * Centrality(i) \tag{14}$$

We have considered several measurements for the core nodes, such as normalized intra-community page-rank centrality and normalized intra-community degree centrality. Both measurements successfully solve the problem of discriminating core nodes. Due to the time complexity, we chose the normalized intra-community degree centrality.

### 3.5 Time Complexity

The time complexity of original LPA is nearly linear, each iteration of propagation has complexity $O(m)$, with m being the number of edges. In our proposed algorithm, we spend extra time to calculate the normalized edge density of all communities during the balance propagation, calculate the intra-community degree centrality of all nodes during the converging propagation, and calculate the normalized loyalty of all nodes during the whole propagation.

When node $i$ updating its label, the number of edges in community $c_i$, the number of neighbors in same community can be updated with time complexity $O(1)$. Then both the edge density of community $c_i$, and the loyalty of node $i$ can be calculated with time complexity $O(1)$ due to the formula 2 and formula 4. At the end of each iteration, the normalized loyalty of all nodes can be calculated with time complexity $O(m)$, the normalized edge density of all communities can be calculated with time complexity $O(|c|)$, the intra-community degree centrality of all nodes can be calculated with time complexity $O(n)$.

If the changed label in one iteration is defined as $d$, as a result, each iteration of our proposed algorithm have time complexity $O(m)$, derived by $O(m) + 4d*O(1) + O(m) + O(|c|) + O(n) + O(n)$, in which $d<n$, $|c|<n$, and $n<m$.

In conclusion, each iteration of roLPA have time complexity $O(m)$, the roLPA algorithm exhibit nearly linear complexity $O(km)$, in which $k$ is the iteration times.

### 3.6 Algorithm

Input: Graph $G(V, E)$, hop attenuation ratio $\delta$
Output: Communities $C$
for $i \in V$ do
   $c_i \leftarrow l_i$
   $c_i' \leftarrow c_i$
   Loyalty($i$) ← 1
   score($v$) ← 1
   $N_i \leftarrow \{j \mid (i,j) \in E \}$
end for
All nodes are ordered in a descending constraint score sequence and stored in queue $V'$;
$V' \leftarrow$ argsort(Constraint($i$))
while not stopping criterion do
   iteration $t \leftarrow t+1$
   for each $i \in V'$ do

```
    if not balance propagation stopping criterion then
        ci ← argmax_i(1 + Density_norm(c_i) - Loyalty_norm(i))
    else
        c_i ← argmax_i(1 + Loyalty_norm(i) * Centrality(i))
    end if
    if c_i ≠ c_i' then
        update Density(c_i)
        update Density(c_i')
        score(i) = max_{j∈N_i}(score(j)) − δ
    end if
  end for
  for i∈V do
    update Loyalty_norm(i)
    update Centrality(i)
  end for
  for c∈C do
    update Density(c)
    update Density_norm(c)
  end for
end while
return C
```

## 4  Experiments and Evaluation

In this section, we present results of empirical evaluation for the proposed algorithm. We first compared algorithms on synthetic networks, namely, GN benchmark networks and LFR benchmark networks. Then we analyzed the proposed algorithm on several real-world networks with community structure, all of which are commonly employed in the community detection literature.

### 4.1 Evaluation Criteria

We use ratio of relative normalized mutual information (rrNMI), and modularity as the evaluation criteria for synthetic and real-world networks respectively.

**I. Ratio of Relative Normalized Mutual Information**

For comparing the identified community detection result and the ground truth, a commonly used measure is normalized mutual information (NMI). The NMI measure the amount of mutual information between two divisions, based on information theory [20]. It's defined as:

$$NMI(X \mid Y) = \left(-2 \times \sum_{i=1}^{|X|} \sum_{j=1}^{|Y|} |X_i \cap Y_i| \times log\left(\frac{(n \times |X_i \cap Y_i|)}{(|X_i| \times |Y_i|)}\right)\right) \times$$

$$\left(\sum_{i=1}^{|X|} |X_i| log\left(\frac{|C_i|}{n}\right) + \sum_{j=1}^{|Y|} |Y_j| log\left(\frac{|C_j|}{n}\right)\right)^{-1} \quad (15)$$

where $n$ represents the number of nodes in the network, $X$ represents a community detection result generated by the evaluated algorithm, and $Y$ represents the ground truth community structure.

As shown by Zhang et al. [21], NMI is seriously affected by systematic error due to finite size of networks, and may give wrong estimate of performance of algorithms in some cases. They proposed relative normalized mutual information(rNMI) to fix this systematic bias. rNMI is defined as follows:

$$rNMI(A, B) = NMI(A, B) - \langle NMI(A, C) \rangle \qquad (16)$$

where $A$ is the planted partition, $B$ is partition obtained by algorithm, $C$ is a random partition with the same group-size distribution as partition $B$, and $\langle NMI(A, C) \rangle$ is the expected NMI between the ground truth configuration $A$ and the random partition $C$, averaged over realization of $C$. Zhang et al. proposed ratio of relative normalized mutual information (rrNMI) [22] to reflect the similarity between the obtained partition $B$ and the planted partition $A$, defined as:

$$rrNMI(A, B) = \frac{rNMI(A,B)}{rNMI(A,A)} \qquad (17)$$

such that it is up-bounded by 1 and equals 1 when partition $B$ is identical to partition $A$.

**II. Modularity**

Modularity proposed by Newman and Girvan [23] is the most used quality function as the community detection evaluation criteria of real-world networks. Modularity measures the quality of a partition of a network into communities by comparing essentially the number of links inside a given community with the expected value for a randomized network of the same size and degrees. It's defined as:

$$Q = \frac{1}{2m} \sum_{i,j \in V} \left( A_{ij} - \frac{d_i d_j}{2m} \right) \times \delta(c_i, c_j) \qquad (18)$$

where $m$ represents the number of edges in the network; $A$ is the adjacency matrix of the network, if node $i$ and node $j$ are directly connected, $A_{ij} = 1$; otherwise, $A_{ij} = 0$; $c_i$ and $c_j$, respectively, denote the label of node $i$ and node $j$, if $c_i = c_j$, then $\delta(c_i, c_j) = 1$, else $\delta(c_i, c_j) = 0$.

**4.2 Resolution Limit Test**

We first test the proposed method on resolution limit test benchmark networks, which are composed by cliques with 4 vertices and each clique is linked to the next clique with an edge to form a ring [24]. Both our method and the label propagation algorithm are run 100 times on each network and the average number of discovered communities is shown in Fig 5. The result shows that our method can discover the correct communities every time. However, the average number of communities discovered by LPA is less than the planted, caused by the core-first infection.

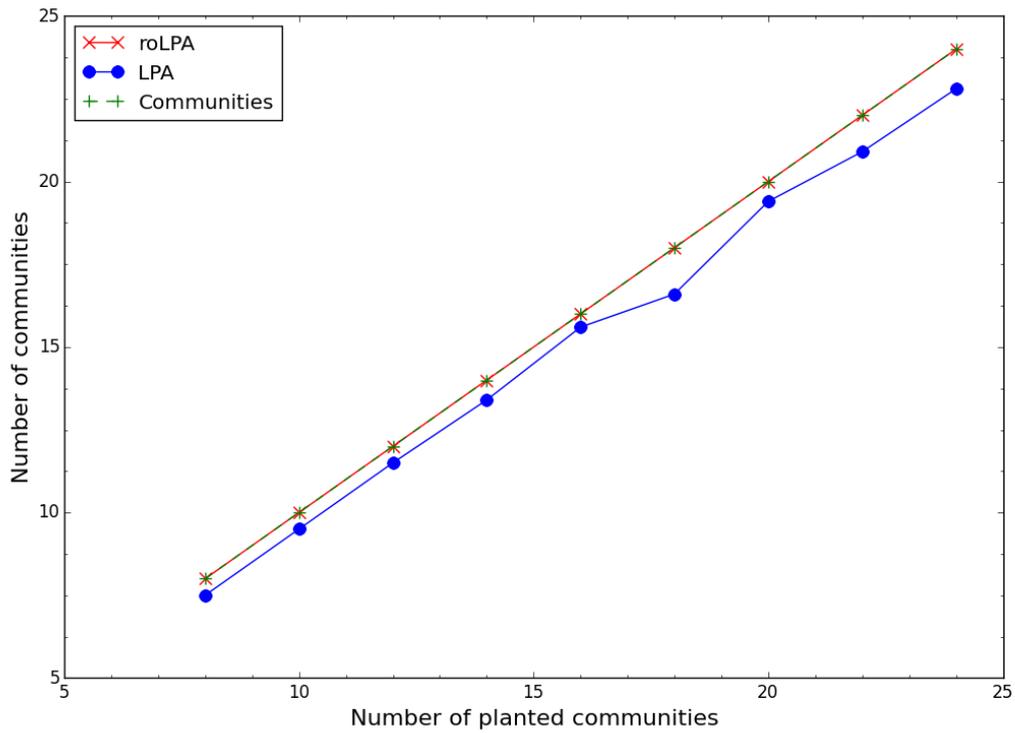

Figure 5. Resolution limit test on benchmark networks.

## 4.3 Synthetic Networks

The first artificial benchmark employed in our experiments is the extended GN benchmark networks [1], which has 128 nodes and is divided into 4 communities. Parameter $\mu$ represents the fraction of the number of links of each node within the community and the degree of the node. If the value of $(1-\mu)$ is large, it means that the community structure of the network is clearer, and vice versa. The result is shown in Fig 7, where y-axis represents the value of rrNMI, and each point in curves is obtained by averaging the values obtained on 100 synthetic networks sampled for each value of the parameter $\mu$. The hop attenuation ratio $\delta$ of LPAD and roLPA is set to 0.10, which is recommended by Leung et al. [5]. The result shown that the other four algorithms perform better than LPA, and our method performs best.

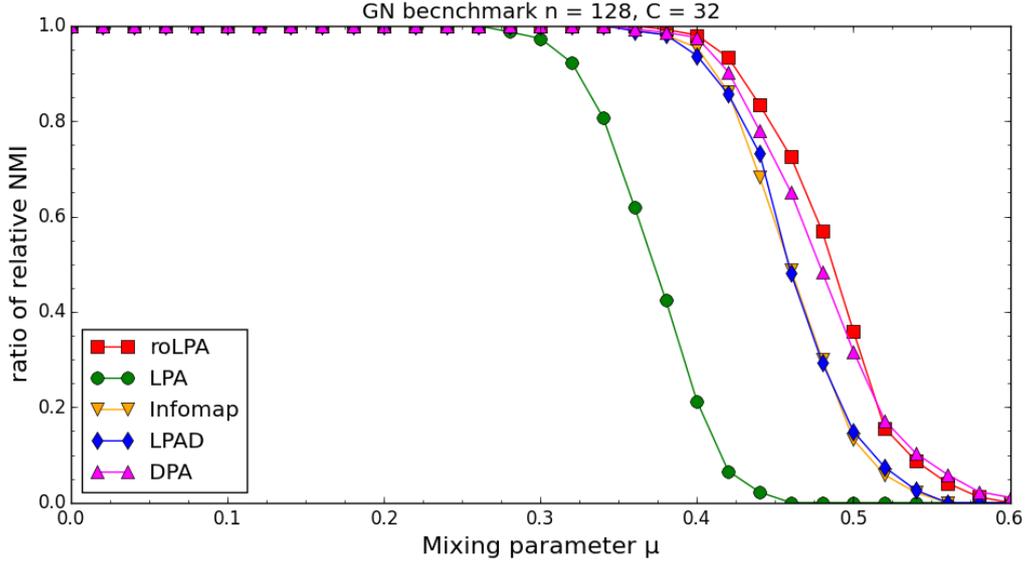

Figure 6. The average results over 100 runs on extended GN benchmark networks.

The second artificial benchmark is the LFR benchmark proposed by Lancichinetti et al. [25]. LFR benchmark exhibits the realistic properties observed in real-world networks, such as the controlled power-law node degree distribution, community size distribution. Hence, it is widely used to generate synthetic network with known ground truth values. Lancichinetti et al. have conducted a thorough empirical analysis of different graph clustering methods including algorithms that are based on modularity maximization and algorithms based on other approaches. In this study, the same LFR networks are generated to test the proposed algorithm.

We firstly generate the LFR benchmark network with given parameters. In the LFR networks, the mixing parameter $\mu$ represents the ratio between the inter-community degree of each vertex and the total degree of the node.

The larger value $\mu$ of the benchmark, the less clear community structure of the networks. We vary the size of the networks ($N$=1000 and $N$=5000) and the size of the communities ($c$=[10, 50] and $c$=[20, 100]) to generate four types of benchmark, which exactly coincide with the benchmarks used in [3].

$N = 1000$, $\langle k \rangle = 20$, $\text{maxk} = 50$, $\tau_1 = 2$, $\tau_2 = 1$, $\text{maxc} = 10$ and $\text{maxc} = 50$,

$N = 1000$, $\langle k \rangle = 20$, $\text{maxk} = 50$, $\tau_1 = 2$, $\tau_2 = 1$, $\text{minc} = 20$ and $\text{maxc} = 100$,

$N = 5000$, $\langle k \rangle = 20$, $\text{maxk} = 50$, $\tau_1 = 2$, $\tau_2 = 1$, $\text{minc} = 10$ and $\text{maxc} = 50$,

$N = 5000$, $\langle k \rangle = 20$, $\text{maxk} = 50$, $\tau_1 = 2$, $\tau_2 = 1$, $\text{minc} = 20$ and $\text{maxc} = 100$.

The experiment is designed for testing the accuracy of our algorithm with various parameter $\mu$ (0.05, 0.1, … , 0.8) on the four types of benchmark.

We compared roLPA with LPA [4], LPAD [5], DPA [11], and Infomap [26]. LPA denotes basic label propagation, LPAD denotes LPA with decreasing hop attenuation, and DPA is proposed by Lovro et al., combining defensive and offensive label propagation. Furthermore, Infomap is a coding theory based algorithm proposed by Rosvall et al., which is the state-of-the-art algorithm analyzed in [3]. The experimental results are displayed in Fig 7, where y-axis represents the value of rrNMI, and each point in curves is obtained by averaging the values obtained on 100 synthetic networks sampled for each value of the mixing parameter. The hop attenuation ratio $\delta$ of LPAD and roLPA is set to 0.10.

For all the algorithms except Infomap, we take the average of 10 repeated trails as the result for each synthetic network for the introduced randomness. In addition, the random partition C in rrNMI is averaged over 100 realizations of C. As we can see in Fig 7, all algorithm works well when $\mu$ is less than 0.5. As the $\mu$ is larger than 0.5, roLPA, DPA, and Infomap both perform much well compared to other algorithms, and our algorithm performs best.

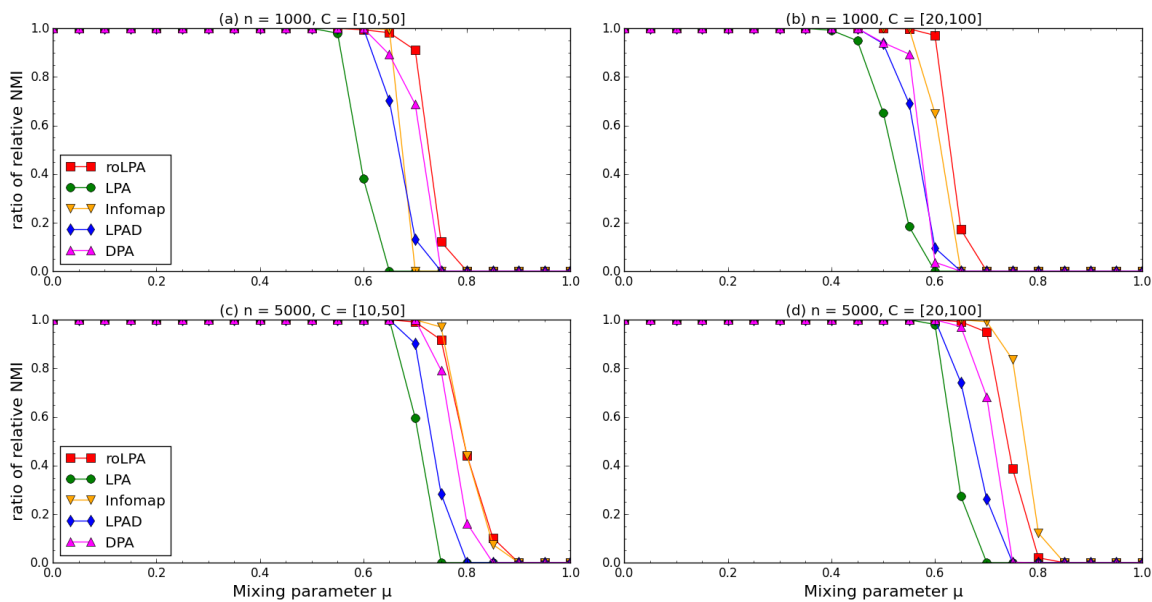

Figure 7. The performance on LFR with respect to the mixing parameter $\mu$.

**4.4 Real-world networks**

In this section, our method was further analyzed on 13 real-world networks involving various types, e.g. social, on-line social, scientist collaboration, citation, electric, metabolic and other networks. Table 1 shows the basic information of these network, the nodes number of which range from tens to millions.

Table 1. Real-world network data. All networks are reduced to the largest component of the original network.

| Network | Description | Nodes | Edges |
| --- | --- | --- | --- |
| karate | Zachary's karate club. [27] | 34 | 78 |
| dolphins | Lusseau's bottlenose dolphins. [28] | 62 | 159 |
| books | Co-purchased political books. [29] | 105 | 441 |
| football | American football league. [30] | 115 | 613 |
| netscience | Scientists working on network theory. [31] | 379 | 914 |
| power | Western U.S. power grid. [32] | 4941 | 6594 |
| cond-mat | Condensed Matter Archive 1999. [33] | 13861 | 44619 |
| hepth | High Energy Physics Archive. [33] | 27400 | 352012 |
| cond-mat-2003 | Condensed Matter Archive 2003. [33] | 27519 | 116181 |
| cond-mat-2005 | Condensed Matter Archive 2005. [33] | 36458 | 171595 |
| dblp | DBLP collaboration network. [34] | 317080 | 1049866 |
| google | Web graph of Google. [35, 36] | 855802 | 4291352 |
| youtube | Youtube online social network. [34] | 1134890 | 2987624 |

We give the results of comparison between 5 different community detection algorithms in Table 2. For each algorithm, we report average modularity derived on the networks. As we can see, roLPA performs as well as the state-of-the-art algorithms on most networks.

Table 2. Average modularities Q for compared community detection algorithms. The results were obtained by running the algorithms from 10 to 10000 times on each network, depending on the size of the network. Delta is set to 0.1 for LPAD and roLPA.

| Network | LPA | LPAD | DPA | Infomap | roLPA |
|---|---|---|---|---|---|
| karare | 0.343 | 0.377 | **0.396** | **0.402** | 0.391 |
| dolphins | 0.478 | 0.485 | 0.503 | **0.528** | **0.520** |
| books | 0.493 | 0.507 | 0.514 | **0.523** | **0.516** |
| football | 0.590 | 0.592 | 0.593 | **0.601** | **0.596** |
| netscience | 0.793 | 0.802 | 0.802 | **0.803** | **0.811** |
| power | 0.803 | 0.834 | **0.892** | 0.816 | **0.875** |
| cond-mat | 0.718 | 0.734 | **0.808** | 0.748 | **0.752** |
| hepth | 0.423 | 0.532 | 0.514 | **0.581** | **0.586** |
| cond-mat-2003 | 0.627 | 0.629 | **0.685** | 0.661 | 0.653 |
| cond-mat-2005 | 0.405 | 0.453 | **0.634** | 0.624 | **0.628** |
| dblp | 0.683 | 0.677 | **0.767** | 0.722 | **0.738** |
| google | 0.828 | 0.829 | **0.966** | 0.836 | **0.842** |
| youtube | **0.573** | 0.569 | 0.544 | 0.556 | **0.595** |

Considering two LPA-based algorithms, namely, roLPA, and DPA, the statistical results of derived communities are different. Particularly speaking, the number of communities generated by roLPA is larger, and the average size of community is smaller. We can infer that DPA is tend to identify the meaningful larger communities, while roLPA is tend to identify more refined communities with more detailed information.

## 5 Conclusion

In this paper, we propose a scalable, accurate and efficient community detection algorithm based on label propagation. We analyze the different dynamic influence of various community-oriented node roles during propagation. With two different propagation phases, the algorithm uses potential community-oriented node role information to confirm the node updating order and the node preference of two phases, by which we eliminate two types of infection and avoid giant community during the propagation. Our algorithm improves the accurate and stable by employing only local measures, and does not require the number of communities to be specified. The proposed algorithm was analyzed on both real-world and artificial benchmark networks with wide range of size and type. The result shows that the accurate of our algorithm is comparable to the state-of-the-art community detection algorithms, and the nearly linear time complexity make the algorithm suitable for community detection in large scale networks.

Despite the advantages of our method compared to previously proposed methods in the literature, there are still some open questions pertaining to explicitly inferring roles. For example, our method lacks a reasonable threshold to exactly terminate the balancing propagation. This challenge is rooted in the fact that there is no exactly definition of a community, accounting for the difficulty in exactly defining bridge nodes. Nevertheless, our approach offers an alternative method for addressing the community detection problem in large-scale complex networks and it is effective and more efficient than existent methods in the literature. Taken together, our approach could motivate further effort towards combining the node role detection to identifying community structures in complex networks.